\documentstyle[psfig,twocolumn,aps,graphicx]{revtex} 
\tighten
\begin{document}

\setlength{\unitlength}{1cm}

\draft

\title{\bf Soliton Lattices in the Incommensurate Spin-Peierls Phase:\\
Local Distortions and Magnetizations}

\author{G\"otz S. Uhrig$^1$, Friedhelm Sch\"onfeld$^1$,
Jean-Paul Boucher$^2$ and Mladen Horvati\'c$^3$}

\address{$^1$ Institut f\"ur Theoretische Physik, Universit\"at zu
  K\"oln, Z\"ulpicher Str. 77, K\"oln 50937, Germany }
\address{$^2$ Laboratoire de Spectrom\'etrie Physique, Universit\'e 
J. Fourier Grenoble I BP 87,\\
 F-38402 Saint-Martin d'H\`eres Cedex, France}
\address{$^3$ Grenoble High Magnetic Field Laboratory, CNRS and MPI-FKF,
 BP 166, 38042 Grenoble Cedex 09, France\\[1mm]
  {\rm(\today)} }

\maketitle

\begin{abstract}
It is shown that nonadiabatic fluctuations of the soliton lattice
in the spin-Peierls system CuGeO$_3$
lead to an important reduction of the NMR line widths. 
These fluctuations are the zero-point motion of the massless phasonic
excitations.
Furthermore, we show that the discrepancy of X-ray and NMR soliton widths
can be understood as the difference between a distortive and a magnetic
width. Their ratio is controlled by the frustration of the spin system.
By this work, theoretical and experimental results can be reconciled in two 
important points. 
\end{abstract}


\narrowtext
\section{Introduction}

Already 20 years ago, the physics around the spin-Peierls transition
fascinated many researchers (for a review, see Ref. \cite{bray83}).
In particular the incommensurably modulated I phase attracted much
interest 
(e.g. \cite{brazo80a,brazo80b,nakan80,nakan81,merts81,horov81,horov87,buzdi83a,buzdi83b,fujit84,fujit88,hijma85}).
Yet detailed experimental investigations of the nature of
this phase were not possible at that time.
The first spin-Peierls transition in an inorganic compound, CuGeO$_3$, was 
found only five years ago \cite{hase93a}. This made a multitude of
experimental investigations possible (for a review, see Ref. \cite{bouch96}).

In particular, direct X-ray experiments in the I phase
were performed by Kiryukhin and Keimer
 which permitted for the first time to detect the
incommensurability of the distortion in $k$-space \cite{kiryu95}.
 Even more, it was
possible  to look at the structure of the soliton lattice modulation by
measuring the intensity of the third harmonic \cite{kiryu96a,kiryu96b}.

On the other side, Fagot-Revurat et al.\cite{fagot96}
were able to measure the distribution of local magnetizations
in CuGeO$_3$ in a beautiful NMR experiment. In a refined version
it was now possible to deduce from such results the shape and the
amplitude of the magnetic part of a soliton \cite{horva98a,horva99}.
Three discrepancies to the conventional theories became apparent.
The first concerns the amplitudes of the local 
magnetizations which is experimentally found to be much lower
(factor 4 to 6) than predicted. Second, the X-ray soliton width 
($13.6\pm0.3$) is appreciably larger than the NMR soliton width
ranging from 6 to 10.
Third, the widths are all larger than the ones theoretically predicted.

In the present work we will solve the first two discrepancies and
argue with Zang et al. \cite{zang97} that the remaining problems
are connected to the neglect of interchain couplings.

To fix the diction let us state that 
we use the term soliton for the combination of a zero in the modulated
distortion {\em and} the concomitant localized, bound spinon 
\cite{nakan80,nakan81,uhrig99a}. The distortive
soliton width is the width of the kink-like zero of the modulated 
distortions. The magnetic soliton width is the spatial width of the local
magnetizations \cite{nakan80,nakan81,schon98,uhrig99a}. The incommensurate
modulation in the I phase is viewed as an equidistant array (lattice)
of solitons.

The paper is set up as follows.
In section II we discuss the fact that the NMR line widths
are much smaller than the theoretical ones and give an explanation for it
in terms of the zero-point motion of the soliton lattice.
Direct numerical calculations based on DMRG are shown in section III.
They permit to address the second main point, namely the discrepancy
between the soliton widths as measured by X-ray and by NMR.
In section IV a detailed comparison to recent experimental data 
\cite{horva98a,horva99} will be presented. The concluding section
contains a discussion of the open questions, namely the role of interchain
couplings, and a summary of our results.

\section{Averaging due to phasons}

In Ref. \cite{fagot96} the results were interpreted by fitting them to a 
Hartree-Fock theory of Fujita and Machida \cite{fujit88}.
Fujita and Machida  did not take into account that the expectation values
which occur in the Hartree-Fock self-consistency problem 
become non-uniform in a non-uniform phase \cite{uhrig98a}. Thus they missed 
the important point that the local magnetizations $ m_i:=\langle
S^z_i \rangle$ are so strongly alternating that they are even antiparallel
on every second site to the applied external magnetic field.
The amplitude of the alternating component is strongly enhanced
compared to the XY model
as was predicted in a number of investigations 
\cite{hijma85,uhrig98a,feigu97,forst98,uhrig99a}.

Theoretically, there is no doubt that the {\em spin-isotropic} model
has to be used to describe cuprate systems. 
Experimentally, however, the amplitude of an XY model fits much better
than the enhanced amplitude of the isotropic XYZ model.
The discrepancy can be explained by the fluctuations of soliton lattice
which are induced by the presence of the so-called phasons \cite{uhrig98a}.

The importance of zero-point motion of the {\em crystal lattice} 
in Peierls systems was already noted by McKenzie and Wilkins
\cite{mcken92}. In the present work we will focus on the zero-point motion
of the phasons. The phasons are very similar to phonons.
If the deviation from commensurability 
of the wave vector characterizing the modulation $d:=|q-\pi|$ 
is small (low soliton concentration) and the soliton width $\xi$ is
large a continuum approach can be used
\cite{brazo80a,brazo80b,nakan80,nakan81,merts81,horov81,horov87,buzdi83a,buzdi83b,fujit84,fujit88,hijma85}.
In this approach the discreteness of the underlying lattice
does not appear. Thus the continuum Hamiltonian is invariant under
continuous translations along the chains. The incommensurate
modulation breaks this continuous symmetry giving rise to Goldstone
bosons, the so-called phasons \cite{bhatt98}. They refer to oscillations of
the solitons about the equilibrium positions in their lattice
just as phonons refer to oscillations of the atoms about the equilibrium
positions in the crystal lattice. There is one important difference
between phonons and phasons. There exist in general 
three phonon branches (2 transversal, 1 longitudinal) corresponding
to the three spatial dimensions into which an atom can be moved.
But there is only one phason branch since the modulation can only
be moved along the chains. Note that this does {\em not}
concern the fact that the phasons have a nondegenerate 
 dispersion $\omega(\vec{k})$ which depends on a
three dimensional vector $\vec{k}$. Like phonons
the dispersion is linear in $\vec{k}$ for small values of $\vec{k}$,
i.e. 
\begin{equation}
\label{omegabhatt}
\omega^2 = (c_x k_x^2 + c_yk_y^2+2|c_z|k_z^2)/\rho
\end{equation}
 in the notation of Ref. \cite{bhatt98}. Hence,
the phasons give rise to a $T^3$ contribution in the specific heat
\cite{bhatt98} which is indeed experimentally observed \cite{loren96}.

Before we proceed further we  discuss briefly the
effect of pinning. First we like to emphasize that there are two possible
sources of pinning. 
The first one is pinning to the discrete lattice structure.
The second is pinning to defects.

The first mechanism enters since the continuous translational invariance
along the chains is given only in the continuum treatment
which represents a certain approximation. So the phasons are only
{\em quasi} Goldstone bosons
 of a {\em quasi} continuous symmetry breaking. Yet treating the
incommensurate modulations as continuously translational invariant,
i.e. shifting them without energy cost,
is an excellent approximation if the solitonic width $\xi$ is not
too small. In the course of our previous calculations \cite{schon98}
we noted that the energy difference between a modulation with the
zero {\em on} a site and a modulation with the zero {\em between}
two sites is of the order $J\exp(-C\xi)$ where $C$ is some constant of the
order of the inverse lattice constant $c^{-1}$ in chain direction.
Hence, for $\xi\approx 10c$ this
energy difference becomes negligibly small. 

From the experimental point
of view we come to the same conclusion. If there were  a pinning of
the modulation to the lattice structure the modulation would be
commensurate with a period $Lc$ where $L$ is an integer. 
This would imply that at most $L$ discrete local magnetization values $m_i$
occur. The experimental resolution, however, is sufficiently high to exclude
this scenario since {\em not} a number of isolated peaks
but a continuous distribution is observed in the NMR response 
 \cite{horva98a,horva99}.

The second mechanism is pinning to defects which break the translational
invariance. Such an effect is certainly present but it is negligibly small
in the pure samples. To obtain an estimate we argue that the defect
concentration $x$ corresponds to a typical distance between two defects
of $l=c/x$ where $c$ is the lattice constant in chain direction. This means
that phasons with a wave vector below $k_z\approx 2\pi/l =2\pi x/c$ do not
exist. This can be viewed as the effect of a gap $\Delta_{\rm pin}$ 
induced by defect pinning. From eq. (\ref{omegabhatt}) we obtain the
estimate 
\begin{eqnarray}
\Delta_{\rm pin} &=& \sqrt{\frac{2 c_z}{\rho}}\frac{2\pi x}{c}\\
&=&  \Delta_{\rm trip} 2\pi\sqrt{2} \frac{\xi_{0z}}{c} x\\
&=&  500 \mbox{K}\;  x
\end{eqnarray}
where we used $c_z^2=u_0\xi^2_{0z}$,
 $\Delta_{\rm trip}=\hbar\sqrt{u_0/\rho}=24$K, 
$\xi_{0z} = 0.69$nm \cite{bhatt98} and  $c=0.294$nm \cite{brade96a}.
An upper bound for the defect concentration $x$ in the pure
samples investigated is $10^{-3}$. So the defect pinning gap
$\Delta_{\rm pin}$ is lower than $0.5$K. We conclude that defect pinning
will become important only below $T\approx 0.5$K and does not need
to be considered here.

The soliton comprises a zero of the modulated distortions and a spinon bound
to this zero \cite{uhrig99a}. If there is a zero-point motion of
the phasons this implies a certain motion of both the lattice
distortions and the magnetic structure. It is plausible to 
assume that a certain {\em averaging} occurs which reduces the
amplitude of the alternating magnetizations. This idea was first
introduced in Ref. \cite{uhrig98a} to explain the difference between
observed and computed magnetization pattern.
(Note that Kiryukhin et al. \cite{kiryu96b} discussed a certain phasonic
averaging linked to {\em defects}. Our approach does not rely on defects.)
 Here we present the detailed
calculation and further estimates.

Let us assume that the local magnetizations $m_i$ can be described 
by two smoothly varying functions $a(r)$ and $u(r)$ which
provide the alternating and the non-alternating component, respectively
\begin{equation}
\label{outset1}
m_i = a({r}_i)\cos(\pi {r}_i) + u({r}_i)
\end{equation}
where we set the lattice constant to unity and $r_i$ denotes the
component along the chains. The continuum approach
results \cite{fujit88,zang97} are in fact of the form (\ref{outset1}).
Eq. (\ref{outset1}) is the adiabatic result describing the completely
static situation without phasons. Let us introduce now the
phase variable $\hat\Theta(\vec{r}_i)$ where we use the hat to indicate
that it is an operator as is the position of a harmonic oscillator.
Thus (\ref{outset1}) becomes
\begin{equation}
\label{outset2}
\hat m_i = a({r}_i)\cos(\pi 
{r}_i+\hat\Theta(\vec{r}_i)) + u({r}_i)\ .
\end{equation}
In principle, the shift $\hat\Theta(\vec{r}_i)$ has to be inserted in
the functions $a({r}_i)$ and $u({r}_i)$ as well. But these functions
are slowly varying so that the influence of the shift on them is 
negligible.
Assuming furthermore that the NMR experiments measure on a relatively
long time scale we conclude that the local magnetization $m_i^{\rm exp}$
seen in experiment is simply the expectation value 
\begin{equation}
\label{outset3}
m_i^{\rm exp} = \langle \hat m_i\rangle
\end{equation}
In the harmonic approximation the phase
fluctuations are interactionless bosons and the operators
$\hat\Theta(\vec{r}_i)$ are linear combinations of the bosonic creation and
annihilation operators. Then it is straightforward to compute
the expectation value of the cosine
\begin{mathletters}
\label{outset4}
\begin{eqnarray}
\label{cosine}
\langle \cos(\pi {r}_i+\hat\Theta(\vec{r}_i))\rangle &=&
\cos( \pi {r}_i) \langle \cos(\hat\Theta(\vec{r}_i))\rangle\\
&=&
\cos( \pi {r}_i) \exp(-\langle\hat\Theta^2(\vec{r}_i)\rangle/2)\ .
\end{eqnarray}
\end{mathletters}
Since the dominant fluctuations are those at long wave lengths
the dependence of $\langle\hat\Theta^2(\vec{r}_i)\rangle$ on the site index
$i$ should not be important. Hence we introduce
\begin{equation}
\label{outset5}
\gamma' := \exp\left(-\frac{1}{2N}\sum_i
\langle\hat\Theta^2(\vec{r}_i)\rangle
\right) < 1
\end{equation}
where $N$ denotes the number of sites in one chain.
The reduction factor $\gamma'$ is similar to  a Debye-Waller
factor which accounts for the non-vanishing atomic motion due to phonons.
 It reduces the amplitude of the alternating component only.
From eqs. (\ref{outset2},\ref{outset3},\ref{outset4},\ref{outset5}) we find
\begin{equation}
\label{result0}
m_i^{\rm exp} = \gamma' a(r_i)\cos(\pi r_i) +u(r_i)
\end{equation}
where the essential amendment compared to (\ref{outset1}) is the 
reduction of the alternating component by $\gamma'$.
It is plausible to explain the discrepancy between experimental
 and adiabatic theoretical amplitude by the zero-point motion of the
phasons which leads to a finite value of $\langle\hat\Theta^2(\vec{r}_i)\rangle$ and hence to $\gamma'<1$. We will present further support
for this idea in section IV.

Before turning to estimates for $\gamma'$ we point out how one can take
the reduction $\gamma'$ into account if the result of the adiabatic
calculation is {\em not} given in the form (\ref{outset1}) but as a set of
discrete values $\{m_i\}$. This is the case for any direct 
adiabatic numerical treatment (see, e.g., section III)
 which does not use the continuum approach.
Then one has to deduce in a first step estimates for the slowly varying
functions $a({r}_i)$ and $u({r}_i)$ from the $m_i$. The most natural way
to do this is  by taking local averages
\begin{mathletters}
\begin{eqnarray}
a({r}_i) &=& m_i/2-(m_{i-1}+m_{i+1})/4\\
u({r}_i) &=& m_i/2+(m_{i-1}+m_{i+1})/4\ .
\end{eqnarray}
\end{mathletters}
After application of $\gamma'$ to $u({r}_i)$ as in (\ref{result0}) one obtains
\begin{mathletters}
\begin{eqnarray}
m^{\rm exp}_i &=& \gamma' a({r}_i) + u({r}_i) \\
 &=& (1+\gamma')m_i/2 +(1-\gamma')(m_{i-1}+m_{i+1})/4\\
&=& (1-2\gamma)m_i +\gamma(m_{i-1}+m_{i+1})
\label{former}
\end{eqnarray}
\end{mathletters}
with $\gamma=(1-\gamma')/4$. Eq. (\ref{former}) is at the basis
 of the averaging of adjacent sites which we used previously \cite{uhrig98a}.
F\"orster {\it et al.} even average completely over two adjacent sites
$m_i \to (m_i+m_{i+1})/2$ \cite{forst98}.

Now we turn to the calculation of $\gamma'$. 
Expressing the expectation
value in the exponent in momentum space yields
\begin{mathletters}
\begin{eqnarray}
&&\frac{1}{N}\sum_j\langle \hat\Theta^2(\vec{r}_j)\rangle
=\frac{1}{N}\sum_{\vec{k}} \langle \hat\Theta^\dagger(\vec{k})
\hat\Theta(\vec{k})\rangle\\
\label{result1}
&&=\frac{\hbar}{2M N}\sum_{\vec{k}}\frac{1}{\omega(\vec{k})}
\left( 1 + \frac{2}{\exp(\hbar\omega/k_{\rm B}T)-1}\right)
\end{eqnarray}
\end{mathletters}
where we used $\hat\Theta(\vec{k})=\sqrt{\hbar/(2M\omega(\vec{k}))}
(\hat a_{\vec{k}} + \hat a^\dagger_{\vec{k}})$ with the mass $M$.
The first term in the bracket in (\ref{result1}) stands for the
zero-point motion since it survives even for $T\to 0$. 
The second term in the bracket  is the bosonic occupation number
at finite temperature.  Rearranging the exponentials yields
\begin{mathletters}
\label{W1}
\begin{eqnarray}
\gamma' &=& \exp(-D/2)\\
D  &=& \frac{\hbar v}{2M}\int\frac{1}{\omega(\vec{k})} 
\coth\left( \frac{\hbar\omega(\vec{k})}{2k_{\rm B}T}\right)
\frac{d^3k}{(2\pi)^3} 
\label{W2}
\end{eqnarray}
\end{mathletters}
where $v$ stands for the volume per spin site.

For a comparison with experimental data we can focus
on the low-temperature behaviour of (\ref{W1}). 
A close inspection of (\ref{W1}) reveals that 
$D=D_1+D_2+{\cal O}(T^3)$ where $D_1$ is constant
and $D_2$ is of order $T^2$. With the help of the
input from Ref. \cite{bhatt98} $D_1$ and $D_2$ are
determined in the appendix. One obtains
\begin{mathletters}
\begin{eqnarray}
\label{width}
D_1 &=& \frac{(3/\pi)^{2/3}}{2\sqrt{2}} \frac{\Delta_{\rm trip}}{u_0\bar\xi_0 v^{2/3}}\\
D_2 &=& \frac{(k_{\rm B}T)^2}{6\sqrt{2}
\bar\xi_0^3 u_0 \Delta_{\rm trip}}
= \left(\frac{T}{T^*}\right)^2 \label{twidth}
\end{eqnarray}
\end{mathletters}
where $\Delta_{\rm trip}$ is the singlet-triplet gap
and $\bar\xi_0$ and $u_0$ a characteristic length and characteristic 
energy per volume, respectively, defined and given in Ref. \cite{bhatt98}.
The value of $D_1$ for CuGeO$_3$ is $3.71$ and the
characteristic temperature $T^*$  is $16.9$K.

The value of $D_1$ leads to a reduction of the alternating 
component $\gamma' = 0.16$ and 
the parameter $\gamma$ takes the value $0.21$. This is in very good
agreement with the values $0.19$ and $0.20$ which we 
found previously by fitting theory to experiment \cite{uhrig98a}.

\begin{figure}
\begin{picture}(8.2,6.2)
\put(0,-0.2){\includegraphics[width=8cm]{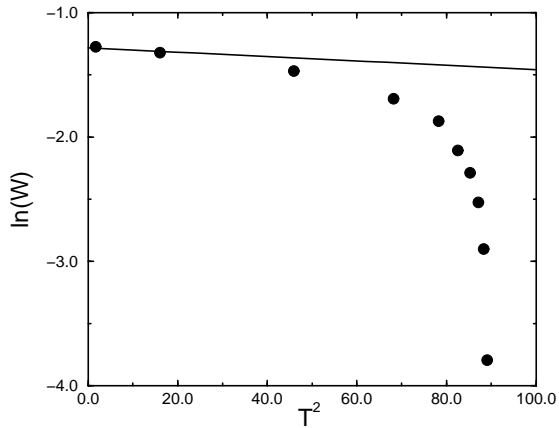}}
% \center{\psfig{width=8cm,file=fig-idea.eps}}
\end{picture}
\caption[]{Solid line: theory from eq. (\ref{twidth});
symbols: experimental data (the error is less than a fifth of the symbol
size).
}
\label{fig-tempdepend}
\end{figure}

Further support is gained from the estimate for $T^*$. 
In Fig. \ref{fig-tempdepend} the temperature dependence of the widths $W$
of the NMR lines are shown which are dominated by the alternating component.
Hence they are expected to be proportional to $\gamma'=\exp(-(T/T^*)^2/2)$
to which they are compared.
The small number of data points at low $T$ 
does not allow for a complete
quantitative test of the theoretical prediction. But the order of 
magnitude of the 
reduction of the line width at low temperatures is the correct one.

If there were an adiabatic treatment of the 
incommensurate phase at finite temperatures this  would yield a line
width which saturates {\em exponentially} at low temperatures due to the
gap in the spin system \cite{klump99a,schon98}. Fig. \ref{fig-tempdepend}
indicates clearly that there is {\em no} exponential saturation of the 
line width for $T\to 0$, but a behaviour $W-W(T=0) \propto -T^2$.
This is a direct evidence for the presence of low-lying, gapless fluctuations.
The quadratic behaviour in $T$ of the decrease corroborates
 the result  (\ref{twidth}) and hence the existence of a three-dimensional
nondegenerate dispersion.

At temperatures closer to the transition the line width is reduced
much more strongly. This is due to the fluctuations in the {\em spin} system
itself which are {\em not} within the scope of the present treatment.

In spite of the crudeness of the present estimates the agreement
gives evidence that the basic idea, phasonic fluctuations leading
to an average of  the local magnetizations, is correct.

\section{Numerical results}
Here we present some numerical results obtained by DMRG.
The calculation treats the phonons adiabatically, i.e.
the local distortions $\delta_i$ are real numbers
which are found by minimizing the ground state energy
of the following  Hamiltonian \cite{feigu97,schon98}
\begin{equation}
\label{hamil}
H = \sum_i\!\! \left[J\left((1+\delta_i) {\bf S}_i{\bf S}_{i+1} + \alpha 
{\bf S}_{i}{\bf S}_{i+2}\right)
+ \frac{K}{2} \delta_i^2  -h S^z_i\right]  .
\end{equation}
The $\delta_i$ are determined self-consistently \cite{feigu97,schon98}.
Numerically, this is achieved by iteration.
Since it is found that an array of equidistant solitons
represents the energetically most favorable configuration 
\cite{uhrig98a,schon98}
we use as initial distortion $\delta_i \propto \cos(qr_i)$
with $q= \pi+ 2\pi m$, $m$ being the average magnetization
or $\delta_i \propto {\rm sign}(\cos(qr_i))$. After about 10 iterations,
one reaches a stable distortion pattern. This pattern shows no discernible 
differences depending on which start configuration has been used. 

For the calculation of the lowest state with $S=1$ for a given modulation we 
use the finite size algorithm \cite{white92,white93}. 
In each iteration we keep 
 $m=64$ states. Periodic boundary conditions are applied. Keeping $m=128$ 
states leads to a change of the calculated energy of the order of $10^{-5}$. 
For large chain lengths and large dimerizations the energy change due to a
 change of the distance by a few sites between two neighboring
 solitons is very
small. This is due to the exponential localization of the solitons.
So care has to be taken to avoid spurious shifts which hinder the
following fit analysis.

\begin{figure}
\begin{picture}(8.2,7.5)
\put(0,0.3){\includegraphics[width=8cm]{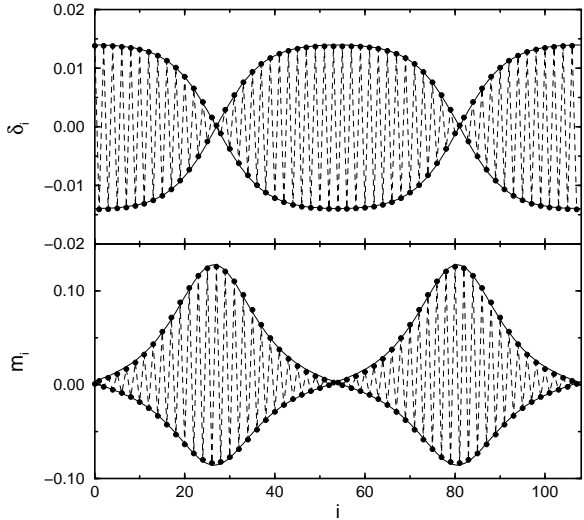}}
\end{picture}
\caption[]{Upper panel: local distortions. Symbols stand for the 
self-consistent  DMRG-result at $K=18J$ and $\alpha=0.35$;
 solid line stems from
 (\protect\ref{fit1b}) with $\delta= 0.014$, $k_{\rm d}=0.959$, and
 $\xi_{\rm d} = 10.5$.
Lower panel: local magnetizations; the solid line stems from
 (\protect\ref{fit1a}) with $W=0.21 $, $R=5.0$, $k_{\rm m}=0.992$, 
and $\xi_{\rm m} = 7.9$.}
\label{fig-k18}
\end{figure}

In Fig. \ref{fig-k18} the local distortions and the local
magnetizations are shown as they are found after the procedure
described above. The solid lines are fits of the following form
\begin{mathletters}
\label{fit1}
\begin{eqnarray}
\nonumber
m_i &=& \frac{W}{2}\Big\{ \frac{1}{R} 
{\rm dn}\left(\frac{r_i}{k_{\rm m}\xi_{\rm m}},k_{\rm m}\right)
+\\\label{fit1a}
&& \hspace*{1cm}+(-1)^i {\rm cn}
\left(\frac{r_i}{k_{\rm m}\xi_{\rm m}},k_{\rm m}\right)\Big\}\\
\delta_i &=& \delta\ 
{\rm sn}\left(\frac{r_i}{k_{\rm d}\xi_{\rm d}},k_{\rm d}\right)\ ,
\label{fit1b}
\end{eqnarray}
\end{mathletters}
where the parameters $W$, $R$, $\delta$, $k_{\rm m}$, and $k_{\rm d}$ are
taken to be the fit parameters.
The periodicity $L$ of the solitons is fixed by the average 
magnetization $m=1/L$ which results in the relation
defining the soliton widths $\xi_{\rm m}$ and $\xi_{\rm d}$
 as a function of $L$ and  $k_{\rm m/d}$
\begin{equation}
\label{m-period}
4 m k_{\rm m/d} K(k_{\rm m/d})\xi_{\rm m/d} = 1 \ .
\end{equation}
Another relation can be deduced from the average of the ${\rm dn}$-function
which is $\pi/(2K(k))$ \cite{abram64} which in the continuum limit
is related to the average magnetization. One obtains
\begin{equation}
\label{m-mittel}
m = \frac{\pi W}{4R K(k_{\rm m})} \ ,
\end{equation}
a relation which was well fulfilled (within 1 to 4\%) by our fit parameters.

The motivation for the equations (\ref{fit1}) is twofold.
First, such fits are used to describe the experimental data
\cite{fagot96,horva98a,horva99}. Second, the continuum approaches provide
results of the above kind 
\cite{brazo80a,brazo80b,horov81,merts81,fujit88,zang97}. The 
continuum results predict that the magnetic and the distortive
parameters are identical 
\begin{equation}
k_{\rm m}=k_{\rm d}
\Leftrightarrow \xi_{\rm m} = \xi_{\rm d}\ .
\end{equation}
The DMRG results can be fitted very well by (\ref{fit1}). In this respect,
the continuum approach works fine. But in other respects it fails.

Taking, for instance, the results for the amplitudes of the
non-alternating and the alternating component \cite{nakan80,nakan81,zang97}
we get for $W$ and $R$
\begin{mathletters}
\label{WR}
\begin{eqnarray}
W &=& \sqrt{\frac{2\Delta_{\rm trip}}{\pi v_{\rm S}}}\\
R &=& k\sqrt{\frac{2\pi v_{\rm S}}{\Delta_{\rm trip}}}
\end{eqnarray}
\end{mathletters}
Inserting some reasonable number for CuGeO$_3$ $\Delta_{\rm trip}=24K, k\approx 1,
v_{\rm S}=\frac{\pi}{2}J(1-1.12\alpha)$ \cite{fledd97}
 with $\alpha=0.35, J=160K$
yields $W=0.32$ and $R=6.3$. So the agreement with the numerical
results presented in Fig. \ref{fig-k18} is not good, but 
the right order of magnitude is reproduced.
Another indication that eqs. (\ref{WR}) have to be extended is the fact that
in a 1D approach 
for $\alpha>\alpha_c=0.2142$ a gap opens and $v_{\rm S}$ is no longer 
well defined.

Similar to the amplitudes, the unique soliton width is given
in the framework of the present continuum theories by
\begin{equation}
\label{solwidth}
\xi = v_{\rm S}/\Delta_{\rm trip}
\end{equation}
which takes the value $6.4$ for the above numbers. Again, this
is too low compared to our numerical results.

The most important discrepancy  is the {\em difference} between the 
magnetic soliton width $\xi_{\rm m}=7.9$ and the distortive 
soliton width  $\xi_{\rm d}=10.5$. Note the ratio 
$\xi_{\rm d}/\xi_{\rm m}=1.33$. This is very interesting
because such a ratio can explain the different experimental findings
for the soliton width. By X-ray measurement \cite{kiryu96a},
$\xi_{\rm d}$ was determined to be $13.6\pm 0.3$ whereas
by NMR $\xi_{\rm m}$ was found to vary between about 10 and 6
with the higher number close to the transition. In fair accordance with
our calculation, the X-ray measurement (susceptible
to the distortion) yields a value  about $1.4$ larger 
than the value obtained by NMR \cite{horva99} (susceptible to the local
magnetizations). We will elucidate this issue further 
at the end of the following section.

\section{Comparison to experimental data}
In this section we attempt a quantitative comparison to
the experimental results obtained recently in high quality
\cite{horva99}. We will do this on
the basis of the Hamiltonian (\ref{hamil}), i.e. a magnetic
one dimensional Hamiltonian with adiabatic phonon treatment.
We use $\alpha=0.35$ and $K=18J$. The value of frustration
results from fits of the temperature dependence of the
susceptibility \cite{riera95,fabri98a}. The value of $K$ is 
then necessary to account for the amount of dimerization
$\delta\approx0.014$ which yields the correct size of the
singlet-triplet gap in CuGeO$_3$ of $\Delta_{\rm trip}/J=24K/160K=0.15$.
This value of $K$ provides also a reasonable estimate for
the critical magnetic field at $T=0$ \cite{loren98}.
One must be aware, however, that the inclusion of higher
dimensional magnetic couplings implies a considerably larger 
dimerization to keep the same gap \cite{uhrig97a}.
A  hint that the actual $K$ value  could be smaller
is provided by the adiabatic analysis of the spin-Peierls
temperature $T_{\rm SP}$. There a value of $K\approx11$
had to be used to reproduce $T_{\rm SP}\approx 14.4$ \cite{klump99a}.
We will discuss the effects of the neglect of the higher dimensional couplings 
in the concluding section.

\begin{figure}
\begin{picture}(8.2,7.5)
\put(0,0.2){\includegraphics[width=8cm]{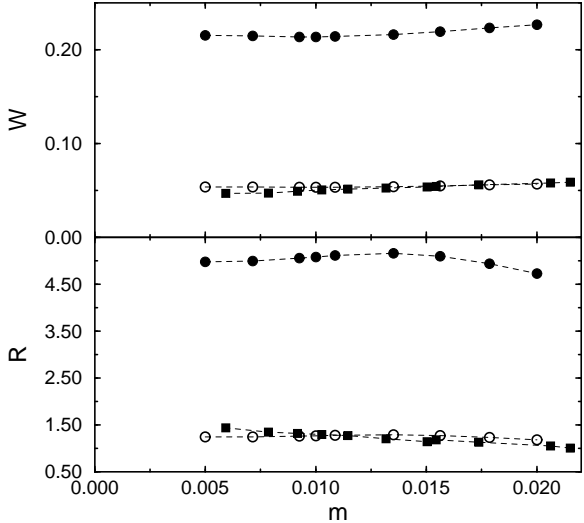}}
\end{picture}
\caption[]{Amplitudes as defined in eq. (\ref{fit1a}). Filled circles:
DMRG calculation for $\alpha=0.35$ and $K=18$ (corresponding to 
$\delta=0.014$); open circles: by $\gamma'=0.25$ renormalized DMRG data.
Filled squares: experimental NMR data after Ref. \cite{horva99}.  }
\label{fig-solitonampli-m}
\end{figure}

In Fig. \ref{fig-solitonampli-m} the amplitudes computed numerically (filled
 circles) as defined in (\ref{fit1a}) are contrasted to
the experimental ones (filled squares; after \cite{horva99}). 
To obtain the computed
values the self-consistently determined patterns are least-square
fitted with the functions given in (\ref{fit1a}). The least square
fits are very good $(\chi^2\approx 10^{-3} - 10^{-4})$, but not excellent.
There are tiny deviations at the magnetic tails of the solitons, see Fig. 
\ref{fig-k18}.
The fits for the lattice distortions on the basis of (\ref{fit1b}) are
better by two orders of magnitude. No deviations are discernible.

For $W$ and $R$, the striking difference between experiment and theory 
can be remedied by using a factor of $\gamma'=0.25$. The 
theoretical results renormalized in this way 
are shown as open circles in Fig. \ref{fig-solitonampli-m}.
The simultaneous reduction of $W$ and of $R$
leaves the homogeneous part $u(r_i)$ of the local magnetization $m_i$
unchanged but reduces the alternating part $a(r_i)$.
This is exactly what we proposed in section II to be the effect of
phasonic zero-point motion. The agreement between renormalized
theory and experiment is good  and provides further support for
the phasonic averaging.

\begin{figure}
\begin{picture}(8.2,7.5)
\put(0,0.2){\includegraphics[width=8cm]{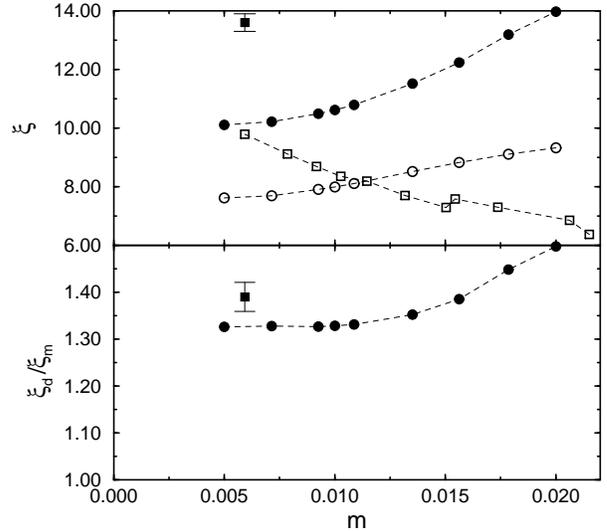}}
\end{picture}
\caption[]{Soliton widths: $\xi_{\rm d}$ distortive;  $\xi_{\rm m}$
magnetic, see eq. (\ref{fit1}). Circles: DMRG results; Open squares:
experimental NMR data after Ref. \cite{horva99}, filled square:
value proposed in Ref. \cite{kiryu96a}. Upper panel: filled symbols are
based  on the distortion; open symbols are
based  on the local magnetizations $m_i$. Lower panel: ratios.}
\label{fig-soliton-m}
\end{figure}

In Fig. \ref{fig-soliton-m} the results for the distortive and the
magnetic soliton widths are plotted. It is obvious that they are
{\em not} the same as was assumed hitherto. Their ratio is to
very good approximation constant for small magnetizations but
grows for larger $m$. This aspect is in qualitative agreement with
the experimental situation where there is also a striking difference
between the X-ray result (distortion) and the NMR result 
(local magnetizations). The fact that the agreement is 
quantitatively not better
can be attributed to several circumstances. The analysis of the X-ray data
\cite{kiryu96a} was done on the assumption of {\em constant} soliton widths
which is not justified. Furthermore, our theoretical
analysis is still based on a one-dimensional model only.
It is also obvious that the qualitative evolution of the magnetic
soliton width is not yet understood. Experimentally, it decreases with
increasing magnetization whereas it increases in our computation.

\begin{figure}
\begin{picture}(8.2,6.2)
\put(0,0){\includegraphics[width=8cm]{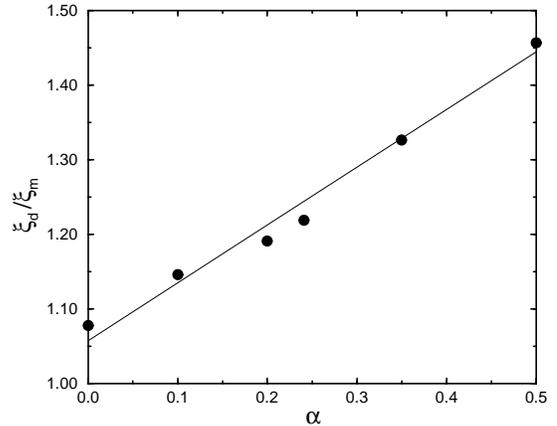}}
\end{picture}
\caption[]{Dependence of the ratio $\xi_{\rm d}/\xi_{\rm m}$ on the frustration
$\alpha$. The values $K$ are chosen such that the $\xi_{\rm d}$ ranges between
5 and 10, i.e. not too small but always way below the sample length of $L=108$.
The values $(\alpha,K)$ are $(0,3), (0.1,4), (0.2,6.2), (0.241,8), (0.35,18),
 (0.5,37)$. Solid line: regression with intercept $1.06$ and slope $0.77$.}
\label{fig-soliton-alpha}
\end{figure}

Based on the lower panel of Fig. \ref{fig-soliton-m} we conclude that
the ratio $\xi_{\rm d}/\xi_{\rm m}$ becomes constant if the solitons
are sufficiently separated. Hence the ratio does not depend on $m$
or $K$ as long as $1/m$ is large enough compared to $\xi_{\rm d}$.
Recall that solitons are exponentially localized objects.
The amplitudes and spatial shapes do not depend 
on the overall energy scale $J$ either.
Thus the only parameter left is the frustration $\alpha$. So we are led
to an analysis of the  ratio $\xi_{\rm d}/\xi_{\rm m}$ as function
of $\alpha$. The results are depicted in Fig. \ref{fig-soliton-alpha}.
It is evident that the frustration is the important control parameter for
the ratio of $\xi_{\rm d}$ and $\xi_{\rm m}$. The dependence seems to be
roughly linear. It should be noted that even at zero frustration the
two soliton widths are not identical. The experimental finding of
a relatively large values of $\xi_{\rm d}/\xi_{\rm m}$ is a strong 
indication for an important frustration in agreement with the
$\chi(T)$ fits \cite{riera95,fabri98a}. A quantitative analysis
appears presently premature since the effects of higher dimensional
couplings are not known yet.

The deviation in the amplitudes as given in eqs. (\ref{WR}) as well as the
non-equality of the soliton widths suggests an extension of the 
existing continuum theories. The fact that the non-equality is strongly
enhanced by the frustration points clearly into the direction that the 
backscattering cosine terms in the sine-Gordon models \cite{nakan81,zang97}
is the origin of this effect. Analysing the self-consistent harmonic
treatment of Nakano and  Fukuyama \cite{nakan80,nakan81} reveals {\em one}
aspect where a difference between the magnetic and the distortive soliton
can appear. This is the so far {\em neglected} spatial dependence of the
renormalizing factors $\exp(-\langle\hat\Theta^2\rangle/2)$
and $\exp(-2\langle\hat\Theta^2\rangle)$, respectively. (Note that this
 operator $\hat\Theta$ is {\em not} related to phasons of distortive origin.
But it stands for gaussian magnetic fluctuations and their formal treatment
bears many analogies to the treatment of phasons.) Further work along these
lines is in progress.

\section{Discussion}
In this work we calculated the effect of phasonic
zero-point motion on the local magnetizations $m_i$. This nonadiabatic
effect can be viewed as virtual oscillations about the
static situation. It leads to a reduction of the alternating
component of the $m_i$. The reduction factor $\gamma'$ as
in eq. (\ref{outset5}) is similar  to a Debye-Waller
factor which is induced by phononic motion.

By the inclusion of the nonadiabatic phason motion
we could explain the so far not understood difference
between experimental NMR line shapes \cite{fagot96,horva98a,horva99}
and previous theoretical predictions 
\cite{hijma85,uhrig98a,feigu97,forst98,uhrig99a}.
Hence the reduction of the NMR line width compared
to the adiabatically predicted one is a strong
evidence for the importance of nonadiabaticity
in CuGeO$_3$. Nonadiabaticity  is also considered to be important for
temperature dependent couplings and induced frustration 
(cf.\cite{uhrig98b,welle98,weiss99b}).

The second main point of our work is the difference
between the distortive and the magnetic soliton width.
To our knowledge, such a difference has
so far not been reported in the literature.
But it helps to understand the differing X-ray and NMR results.
The ratio $\xi_{\rm d}/\xi_{\rm m}$ is controlled by the frustration
$\alpha$. It is fairly close but not equal to unity at zero frustration
and grows roughly linear to almost $1.5$ at the Majumdar-Ghosh point
$\alpha=1/2$. Thus, further evidence for a sizable frustration
in CuGeO$_3$ is provided.

Aiming for quantitative agreement with experiment some questions remain.
Choosing a small dimerization (the corresponding large $K$, respectively), 
which reproduces the correct gap value and critical magnetic field
in a one-dimensional model, leads to amplitudes $W$ and $R$ which
have to be renormalized by about $\gamma'=0.25$ and not by $\gamma'=0.16$
as estimated in section II. Thus these amplitudes seem to be 
too small. The magnetic soliton width $\xi_{\rm m}$, is too low
by about 20\%. 

Choosing, however, a larger dimerization  
(the corresponding small $K$, respectively),
which reproduces the gap value resulting from
averaging the dispersion $\omega(\vec{k})$ perpendicular
to the chains along $k_{\rm b}$ \cite{uhrig97a}, 
leads to larger amplitudes $W$ and $R$. These
larger amplitudes can be nicely reconciled with the
experimental findings by a value very close to $\gamma'=0.16$.
This is essentially what was done in the previous work \cite{uhrig98a}.
The argument in favor of the larger gap is that the neglect of
the interchain coupling would lead to this larger gap \cite{uhrig97a}.
But the corresponding soliton widths would be even lower 
(see eq. (\ref{solwidth})) and hence in worse agreement with
experiment. We conclude from this that it is not just
a question of fine-tuning the constants to achieve
agreement. The purely $d=1$ model is {\em not} sufficient.

The magnetic interchain couplings have to be taken into account
as has become already apparent from the two ways to
choose the relevant gap. The magnetic interchain coupling has the 
two following effects. First, it reduces the observed gap value for a given
dimerization \cite{uhrig98a}.
Thus a larger dimerization (smaller $K$) should be used.
This is consistent with a number of recent investigations
\cite{buchn98,werne98}. The larger gap will lead to larger
amplitudes.
Second, Zang et al. \cite{zang97} argued that the interchain
coupling {\em enhances} the soliton widths which is exactly
what is needed to reconcile theory and experiment.
This enhancement is qualitatively easy to understand since an interchain
coupling will favor the appearance of staggered magnetization.
Regions of staggered magnetization are those were the solitons
 are, see Fig. \ref{fig-k18}. So these regions are extended
which implies an enhancement of the soliton width.
Hence, we come to the conclusion that a larger dimerization (smaller $K$)
plus an appropriate treatment of the interchain couplings should
yield results in quantitative agreement with experiment.
There are also preliminary results indicating that the inclusion of
interchain coupling changes the increase of $\xi_{\rm m}$ as function of
$m$ to a decrease which is what is experimentally seen 
(cf. Fig. \ref{fig-soliton-m}).

Summarizing, two main points are shown in the present work.
The first is the importance of the phasonic fluctuations 
providing a natural explanation for the large amplitude differences
between theory and experiment. The second is the difference between
distortive ($\xi_{\rm d}$) and magnetic ($\xi_{\rm m}$) soliton width
leading to a ratio  $\xi_{\rm d}/\xi_{\rm m}$ between $1.05$ and $1.45$
depending on frustration. Further work to include interchain coupling is
called for.

We gratefully acknowledge useful discussions with C. Berthier and 
D. F\"orster and the support of this work by E. M\"uller-Hartmann.
Part of this work (GSU,FS) was supported by the SFB 341 of the DFG.

%\bibliographystyle{prsty}
%\bibliography{../bibinput/liter10}

\begin{thebibliography}{10}

\bibitem{bray83}
J.~W. Bray, L.~V. Interrante, I.~C. Jacobs, and J.~C. Bonner,  in {\em Extended
  Linear Chain Compounds}, edited by J.~S. Miller (Plenum Press, New York,
  1983), Vol.~3, p.\ 353.

\bibitem{brazo80a}
S.~A. Brazovski\v{\i}, Sov. Phys. JETP {\bf 51},  342  (1980).

\bibitem{brazo80b}
S.~A. Brazovski\v{\i}, S.~A. Gordyunin, and N.~N. Kirova, JETP Lett. {\bf 31},
  456  (1980).

\bibitem{nakan80}
T. Nakano and H. Fukuyama, J. Phys. Soc. Jpn. {\bf 49},  1679  (1980).

\bibitem{nakan81}
T. Nakano and H. Fukuyama, J. Phys. Soc. Jpn. {\bf 50},  2489  (1981).

\bibitem{merts81}
J. Mertsching and H.~J. Fischbeck, Phys. Stat. Sol. (b) {\bf 103},  783
  (1981).

\bibitem{horov81}
B. Horovitz, Phys. Rev. Lett. {\bf 46},  742  (1981).

\bibitem{horov87}
B. Horovitz, Phys. Rev. B {\bf 35},  734  (1987).

\bibitem{buzdi83a}
A.~I. Buzdin and V.~V. Tugushev, Sov. Phys. JETP {\bf 58},  428  (1983).

\bibitem{buzdi83b}
A.~I. Buzdin, M.~L. Kuli\'c, and V.~V. Tugushev, Solid State Commun. {\bf 48},
  483  (1983).

\bibitem{fujit84}
M. Fujita and K. Machida, J. Phys. Soc. Jpn. {\bf 53},  4395  (1984).

\bibitem{fujit88}
M. Fujita and K. Machida, J. Phys. {\bf C21},  5813  (1988).

\bibitem{hijma85}
T.~W. Hijmans, H.~B. Brom, and L.~J. de~Jongh, Phys. Rev. Lett. {\bf 54},  1714
   (1985).

\bibitem{hase93a}
M. Hase, I. Terasaki, and K. Uchinokura, Phys. Rev. Lett. {\bf 70},  3651
  (1993).

\bibitem{bouch96}
J.~P. Boucher and L.~P. Regnault, J. Phys. I France {\bf 6},  1939  (1996).

\bibitem{kiryu95}
V. Kiryukhin and B. Keimer, Phys. Rev. B {\bf 52},  704  (1995).

\bibitem{kiryu96a}
V. Kiryukhin, B. Keimer, J.~P. Hill, and A. Vigliante, Phys. Rev. Lett. {\bf
  76},  4608  (1996).

\bibitem{kiryu96b}
V. Kiryukhin {\it et~al.}, Phys. Rev. B {\bf 54},  7269  (1996).

\bibitem{fagot96}
Y. Fagot-Revurat {\it et~al.}, Phys. Rev. Lett. {\bf 77},  1861  (1996).

\bibitem{horva98a}
M. Horvati\'c {\it et~al.}, Physica {\bf B246-247},  22  (1998).

\bibitem{horva99}
M. Horvati\'c {\it et~al.}, condmat/9812370.

\bibitem{zang97}
J. Zang, S. Chakravarty, and A.~R. Bishop, Phys. Rev. B {\bf 55},  R14705
  (1997).

\bibitem{uhrig99a}
G.~S. Uhrig, F. Sch\"onfeld, M. Laukamp, and E. Dagotto, Eur. Phys. J. B {\bf
  7},  67  (1999).

\bibitem{schon98}
F. Sch\"onfeld, G. Bouzerar, G.~S. Uhrig, and E. M\"uller-Hartmann, Eur. Phys.
  J. B {\bf 5},  521  (1998).

\bibitem{uhrig98a}
G.~S. Uhrig, F. Sch\"onfeld, and J. Boucher, Europhys. Lett. {\bf 41},  431
  (1998).

\bibitem{feigu97}
A.~E. Feiguin, J.~A. Riera, A. Dobry, and H.~A. Ceccatto, Phys. Rev. B {\bf
  56},  14607  (1997).

\bibitem{forst98}
D. F\"orster, Y. Meurdesoif, and B. Malet,  condmat/9802245.

\bibitem{mcken92}
R.~H. McKenzie and J.~W. Wilkins, Phys. Rev. Lett. {\bf 69},  1085  (1992).

\bibitem{bhatt98}
S.~M. Bhattacharjee, T. Nattermann, and C. Ronnewinkel, Phys. Rev. B {\bf 58},
  2658  (1998).

\bibitem{loren96}
T. Lorenz, U. Ammerahl, R. Ziemes, and B. B\"uchner, Phys. Rev. B {\bf 54},
  R15610  (1996).

\bibitem{brade96a}
M. Braden {\it et~al.}, Phys. Rev. B {\bf 54},  1105  (1996).

\bibitem{klump99a}
A. Kl\"umper, R. Raupach, and F. Sch\"onfeld, Phys. Rev. B {\bf 59},  3612
  (1999).

\bibitem{white92}
S.~R. White, Phys. Rev. Lett. {\bf 69},  2863  (1992).

\bibitem{white93}
S.~R. White, Phys. Rev. B {\bf 48},  10345  (1993).

\bibitem{abram64}
M. Abramowitz and I.~A. Stegun, {\em Handbook of Mathematical Functions} (Dover
  Publisher, New York, 1964).

\bibitem{fledd97}
A. Fledderjohann and C. Gros, Europhys. Lett. {\bf 37},  189  (1997).

\bibitem{riera95}
J. Riera and A. Dobry, Phys. Rev. B {\bf 51},  16098  (1995).

\bibitem{fabri98a}
K. Fabricius {\it et~al.}, Phys. Rev. B {\bf 57},  1102  (1998).

\bibitem{loren98}
T. Lorenz {\it et~al.}, Phys. Rev. Lett. {\bf 81},  148  (1998).

\bibitem{uhrig97a}
G.~S. Uhrig, Phys. Rev. Lett. {\bf 79},  163  (1997).

\bibitem{uhrig98b}
G.~S. Uhrig, Phys. Rev. B {\bf 57},  R14004  (1998).

\bibitem{welle98}
G. Wellein, H. Fehske, and A.~P. Kampf, Phys. Rev. Lett. {\bf 81},  3956
  (1998).

\bibitem{weiss99b}
A. Wei\ss{}e, G. Wellein, and H. Fehske,  condmat/9901262.

\bibitem{buchn98}
B. B\"uchner, H. Fehske, A.~P. Kampf, and G. Wellein, condmat/9806022.

\bibitem{werne98}
R. Werner, C. Gros, and M. Braden,  condmat/9810038.

\bibitem{ashcr76}
N.~W. Ashcroft and N.~D. Mermin, {\em Solid State Physics} (Saunders College,
  Philadelphia, 1976).

\end{thebibliography}

\appendix

\section*{}
Based on eq. (\ref{W2}) and Ref. \cite{bhatt98} we present here estimates for
the two leading contributions $D_1$ and $D_2$. Focussing on the
low-lying excitations we adopt  (\ref{omegabhatt})
even though we do not know whether at larger values of $\vec{k}$
the dispersion is still described by (\ref{omegabhatt}).

The mass density $\rho$ is proportional to $M/v$.
If each spin site moved in the same way on shifting
the modulation $M = v \rho$ were reasonable. Since, however,
the modulation does {\em not} have the same amplitude for {\em all}
sites an effective reduction has to be taken into account.
For sinusoidal modulation this factor is 1/2, i.e. $M=v\rho/2$,
 due to the average
value of $\sin^2$. For all other modulations the reduction factor
will be between 1/2 and 1; we choose 1/2 since
 the higher harmonic content of the
 modulation is very small \cite{kiryu96a}.

With $k_x \to k_x\sqrt{\rho/c_x}$, $k_y \to k_y\sqrt{\rho/c_y}$, 
and $k_z \to k_z\sqrt{\rho/|2c_z|}$ as well as
$c_{i}= r_0T_0 \xi^2_{0i}$ and $\bar\xi_0 = 
(\xi_{0x}\xi_{0y}\xi_{0z})^{1/3}$ we obtain
\begin{mathletters}
\begin{eqnarray}
D &=& \frac{\hbar}{\rho} \frac{\rho^{3/2}}{\sqrt{2c_xc_yc_z}}
\int \frac{d^3k}{(2\pi)^3} \frac{1}{k} \coth\left(\hbar k/(2k_{\rm B}T)\right)
\\
 &=& \frac{\hbar}{\rho\sqrt{2}} \left(\frac{\rho}{\bar\xi^2_0r_0T_0}\right)^{3/2}
\int \frac{d^3k}{(2\pi)^3} \frac{1}{k} \coth\left(\frac{\hbar k}{2k_{\rm B}T}\right)
\end{eqnarray}
\end{mathletters}
For the value of $u_0=r_0T_0$ and other values see  Ref. \cite{bhatt98}.
We split $D$ in the zero temperature contribution $D_1$ and
the temperature dependent rest $D_2$ by using $\coth(x/2) = 1+2/(\exp(x)-1)$.

We estimate $D_1$, the zero temperature reduction due to the
zero-point motion of the phasons. It is  difficult to
compute $D_1$ reliably since, in principle, information of {\em all}
 the phasons is required, not only the lowest lying ones. But an estimate 
which determines the order of magnitude is possible
by using an upper cutoff $k_{\rm max}$ as in the Debye model of phonons
\cite{ashcr76}. We start with
\begin{mathletters}
\begin{eqnarray}
D_1 &=& \frac{\hbar}{\rho}\int \frac{d^3k}{(2\pi)^3}\frac{1}{\omega(\vec{k})}\\
&=& \frac{\hbar\sqrt{\rho}}{2\sqrt{2}\pi^2
\left(\bar\xi_0\sqrt{u_0}\right)^3}\int_0^{k_{\rm max}} k dk\\
&=& \frac{\hbar\sqrt{\rho}}{4\sqrt{2}\pi^2
\left(\bar\xi_0\sqrt{u_0}\right)^3} k^2_{\rm max} \ .
\label{w1}
\end{eqnarray}
\end{mathletters}
The value for $k_{\rm max}$ after rescaling 
is  obtained from the
number of possible phasonic excitations 
\begin{mathletters}
\begin{eqnarray}
1 &=& v\int\frac{d^3k}{(2\pi)^3}\\
 &=& \left(\frac{\rho}{\bar\xi_0^2u_0}\right)^{3/2}\frac{v}{2\sqrt{2}\pi^2}
\int_0^{k_{\rm max}}k^2dk \\
\Rightarrow k_{\rm max} &=& \sqrt{\frac{u_0}{\rho}}
\bar\xi_0 \left(\frac{6\sqrt{2}\pi^2}{v}\right)^{1/3}\ .
\label{kmax}
\end{eqnarray}
\end{mathletters}
Eqs. (\ref{w1},\ref{kmax}) and the singlet-triplet gap
$\Delta_{\rm trip}=\hbar\sqrt{u_0/\rho}$ \cite{bhatt98}
 together lead finally to (\ref{width}).
It should be noted that the volume per spin site can be found from
the lattice constants by
$v=abc/2$ with $a=4.79\cdot 10^{-10}m$, $b=8.40\cdot 10^{-10}m$,
and $c=2.94\cdot 10^{-10}m$ \cite{brade96a}. The division by 2 is necessary
since there are two Cu ions in each unit cell.

Next we compute $D_2$ where we are only interested in the leading
temperature dependence. Thus the result depends only on the lowest
lying excitations and the assumption of a dispersion as (\ref{omegabhatt}) is
well justified. We find
\begin{mathletters}
\begin{eqnarray}
D_2 &=& \frac{\hbar}{\sqrt{2}\pi^2\rho}\left( 
\frac{\rho}{\bar\xi^2_0u_0}\right)^{3/2}\!\!\int_0^\infty \!\!\!
\left(\!\exp(\frac{\hbar k}{k_{\rm B}T})-1\!\right)^{-1}\!\!\!\!\!\! kdk\\
&=& \frac{(k_{\rm B}T)^2}{\sqrt{2}\pi^2\hbar \rho} 
\left(\frac{\rho}{\bar\xi^2_0u_0}\right)^{3/2}\!\!
\int_0^\infty\!\!\! \left(\exp(k)-1\right)^{-1}k dk\\
&=& 
\frac{\sqrt{\rho}(k_{\rm B}T)^2}{6\sqrt{2}
\hbar\bar\xi_0^3 u_0^{3/2}}\ =\
\frac{(k_{\rm B}T)^2}{6\sqrt{2}
\bar\xi_0^3 u_0 \Delta_{\rm trip}}\\
&=& \left(\frac{T}{T^*}\right)^2\ .
\end{eqnarray}
\end{mathletters}
With the numbers $\Delta_{\rm trip}=24$K, $u_0 = 650$mJ/cm$^3$ and
$\bar\xi_0$ = 0.31nm the characteristic temperature $T^*$ can be
estimated to be $16.9$ K. 
\end{document}